\documentclass[11pt]{article}
\usepackage{amsfonts,amsmath,amssymb,mathrsfs}
\usepackage[footnotesize]{caption}% 
\usepackage{graphicx}% Include figure files

\newcommand{\RRR}{\mathbb{R}}

\begin{document}
\begin{large}
\title{\bf {Generalization of the Born rule}}
\author{Bruno Galvan \footnote{E-mail: b.galvan@virgilio.it}\\ \small via Melta 16, 38100 Trento, Italy.}
\date{\small \today}
\maketitle
\end{large}
\begin{abstract}
A new formulation of quantum mechanics is proposed based on a new principle that can be considered a generalization of the Born rule. The principle is composed of a mathematical expression and an associated interpretation, and establishes a correlation between the positions of a particle at two different times. Under reasonable conditions for the wave function, this correlation implies that the particles follow quasi-classical trajectories. It is also shown that the Born rule is equivalent to a particular case of the new principle. These features allow the principle to provide a unified explanation of the results of the statistical experiments and of the quasi-classical macroscopic evolution.

There is a strong analogy between the new quantum principle and a probabilistic principle which is necessary to derive empirical predictions from the mathematical formalism of probability theory. This principle is referred to by some authors as Cournot's principle, while other authors use the equivalent notion of typicality. In this paper probability theory and quantum mechanics are formulated in such a way as to explicitly include the two principles and to emphasize the very similar conceptual structure of the two theories.
\end{abstract}
\section{Introduction} \label{introduction}

A reasonably simple and universally accepted way to derive classical mechanics from quantum mechanics does not yet appear to be available. In textbooks the Ehrenfest theorem is often considered as the way to deduce classical mechanics from quantum mechanics, according to the following reasoning: let us suppose that the wave function of a particle is localized in a region which is small at the macroscopic level. Thus a definite macroscopic position can be attributed to the particle, namely the mean value $\langle Q \rangle (t) :=\langle \Psi(t) | Q | \Psi(t) \rangle$. From the Ehrenfest theorem we can deduce that
\begin{equation} 
\frac{d }{dt} \langle Q \rangle=\frac{\langle P \rangle}{m} \; \; \hbox{and} \; \; 
\frac{d }{dt} \langle P \rangle = - \langle \nabla V \rangle \approx - \nabla V (\langle Q \rangle), 
\end{equation}
where $\langle P \rangle$ and $\langle \nabla V \rangle$ are defined analogously to $\langle Q \rangle$, and the potential $V$ is assumed to be approximately constant in the region in which the particle is localized. The problem is that during its evolution a localized wave function may spread over a region which is no longer localized at the macroscopic level. Thus a definite position can no longer be associated with the particle, and the Ehrenfest theorem no longer functions.

The situation in which the wave function spreads over a region which is not localized at the macroscopic level constitutes the infamous {\it measurement problem} of quantum mechanics, and no general consensus exists in the physics community about how to cope with this problem. At least three main different solutions have been proposed: (i) the many words interpretation (MVI), according to which when the wave function splits into macroscopically distinct parts the observers also split correspondently \cite{everett, mwi}. (ii) The dynamical collapse theory of Ghirardi, Rimini and Weber (GRW), according to which a stochastic term added to the unitary evolution of the wave function determines its random collapse and maintains its localization \cite{grw}. (iii) Bohmian mechanics, according to which the particles follow trajectories determined by the wave function through the so-called guidance equation \cite{bohm1, allori}.

In my opinion none of these theories has been universally accepted by the physics community as the true solution.

\vspace{3mm}
In this paper a new formulation of quantum mechanics is proposed to solve this problem. According to this new formulation particles follow definite trajectories, as in Bohmian mechanics, but their structure is defined by a new principle which can be considered as a generalization of the Born rule. The principle is based on the following very natural assumption: suppose that a particle is prepared at a time $t_I$ in such a way that its wave function is split into two impenetrable boxes, and that a time $t_F > t_I$ the boxes are open in order to determine in which of them the particle is contained. The assumption is that the particle was since the time $t_I$ in the box where it is found at the time $t_F$. This assumption is not deducible from the postulates of standard quantum mechanics. Nevertheless, from the physical point of view it is very reasonable, because the boxes are impenetrable and the particle cannot jump between them. The first consequence of this assumption is that the wave function is no longer the most complete description of the particle, because in the time interval $[t_I, t_F]$ the particle is in one of the two boxes, but this information is not present in the wave function. It is therefore natural to assume that at every time the particle has a definite position. The new principle then establishes in a formal way that if the wave function of a particle is split into non-overlapping parts, the particle stays inside the support of one of the parts, without jumping to the others.

The new principles establishes a correlation between the positions of the particles at two different times. Under reasonable conditions for the wave function, this correlation implies that the particles follow trajectories which have a quasi-classical structure on a macroscopic scale. It will also be shown that the Born rule is a particular case of this principle. The new principle therefore has the attractive feature of explaining both the results of the statistical experiments and the quasi-classical macroscopic evolution in an economic and unified way.

There is another attractive feature of the principle, which requires some explanation. The connection of the mathematical formalism of probability theory with the empirical world is based on a principle establishing that we can be empirically certain that an event with probability very close to 1 will happen in a single trial of an experiment. Sometimes this principle is referred to as Cournot's principle; other authors express the principle in terms of typicality. See section \ref{cournot} and the references cited there. The new quantum principle can be considered as the quantum version of Cournot's principle, and actually reduces to it in a particular case. For this reason it will be referred to as the {\it quantum Cournot principle}. Moreover, in the new formulation of quantum mechanics a quantum system can be represented by a mathematical structure which is very similar to a probability space, in the same way in which a statistical experiment is represented by a probability space. The quantum Cournot principle therefore has the merit of showing a strong analogy between two theories, probability theory and quantum mechanics, which are often considered incompatible.

This paper is my fourth paper on this subject \cite{galvan1, galvan2, galvan3}. It is however self-contained, and in a complete and coherent form presents many concepts that were not yet well understood in the previous papers. The paper is structured as follows: Section \ref{some} introduces some preliminary notations. Section \ref{cournot} illustrates the probabilistic Cournot principle and proposes a formulation of probability theory which explicitly includes the principle. Section \ref{quantum} introduces the quantum Cournot principle and proposes the new formulation of quantum mechanics deriving from this principle. Section \ref{consistency} discusses the consistency of the quantum Cournot principle. Section \ref{statistical} shows that the quantum Cournot principle incorporates the empirical predictions of the Born rule. Section \ref{multiple} analyses the nature of the multiple time correlations predicted by the quantum Cournot principle. Section \ref{quasi} then shows how the quantum Cournot's principle defines trajectories with a quasi-classical structure. Section \ref{analysis} discusses some classical quantum experiments in the light of the new formulation. Section \ref{retrodictions} discusses the empirical verifiability of the quantum Cournot principle. Section \ref{summary} summarizes the paper.

\section{Some preliminary notations} \label{some}

A probability space $\cal P$ is the triple $(\Omega, {\cal F}, P)$, where $\Omega$ is the sample space, ${\cal F}$ is the $\sigma$-algebra of the events and $P$ is the probability measure. Given two probability spaces ${\cal P}_1=(\Omega_1, {\cal F}_1, P_1)$ and ${\cal P}_2=(\Omega_2, {\cal F}_2, P_2)$, the symbol ${\cal P}_1 \times {\cal P}_2$ will denote the probability space $(\Omega_1 \times \Omega_2, {\cal F}_1 \times {\cal F}_2, P_1 \times P_2)$, where ${\cal F}_1 \times {\cal F}_2$ is the $\sigma$-algebra generated by the Cartesian product of ${\cal F}_1$ and ${\cal F}_2$ and $P_1 \times P_2$ is the unique measure on $\Omega_1 \times \Omega_2$ such that $P_1 \times P_2(\Delta_1 \times \Delta_2)=P_1(\Delta_1) P_2(\Delta_2)$.

$X$ denotes the configuration space of a physical system, and $\cal B$  an appropriate $\sigma$-algebra of subsets of $X$. For example, for an N-particle system $X=\RRR^{3N}$ and $\cal B$ is the Borel $\sigma$-algebra. The evolution of a system will always be considered during a time interval $T=[t_I, t_F]$, which will be assumed to include the origin $t=0$. The set of all the trajectories $\lambda$ from $T$ to $X$ is denoted by $X^T$. Given $\Delta \in {\cal B}$ and $t \in T$, the pair $(t, \Delta)$ denotes the set $\{\lambda \in X^T: \lambda(t) \in \Delta\}$. Sets of this type are referred to as {\it s-sets}, which is an abbreviated name for single-time cylinder sets. The class of the s-sets will be denoted by $\cal S$, and the s-sets will also be denoted by $S, S_1, \ldots $. The set $\cal S$ is not a $\sigma$-algebra; note however that $S_1 \cap S_2 = \emptyset$ implies that $S_1 \cup S_2 \in {\cal S}$.

The $\sigma$-algebra of subsets of $X^T$ generated by $\cal S$ is denoted by $\sigma({\cal S})$; it is equal to the $\sigma$-algebra generated by the cylinder sets of $X^T$. By endowing $\sigma({\cal S})$ with a probability measure $P$ we obtain the {\it canonical stochastic process} $(X^T, \sigma({\cal S}), P)$.

$\cal H$ denotes the Hilbert space of a quantum system, $U(t)$ the unitary time evolution operator, $\Psi_0$ the normalized state of the quantum system at the time $t=0$ and $\Psi(t) =U(t)\Psi_0$ the state of the system at the time $t$. $E$ denotes the usual spatial projection valued measure $E:{\cal B} \rightarrow \hbox{Pr}({\cal H})$, and $\hat{\Psi}$ the vector-valued set function $\hat{\Psi}:{\cal S} \rightarrow {\cal H}$ defined by
\begin{equation} \label{yu}
\hat{\Psi}[(t, \Delta)]:=U^{-1}(t)E(\Delta)U(t)\Psi_0.
\end{equation}
The hat distinguishes the set function $\hat{\Psi}$ from the wave function $\Psi(t)$. Note that $\hat{\Psi}$ is $\sigma$-additive. 

The triple ${\cal Q}=(X^T, {\cal S}, \hat{\Psi})$ is referred to as a {\it quantum process}. Given two quantum processes ${\cal Q}_1 = (X_1^T, {\cal S}_1, \hat{\Psi}_1)$ and ${\cal Q}_2=(X_2^T, {\cal S}_2, \hat{\Psi}_2)$, the symbol ${\cal Q}_1 \times {\cal Q}_2$ denotes the quantum process $((X_1 \times X_2)^T, {\cal S}_1 \times {\cal S}_2, \hat{\Psi}_1 \otimes \hat{\Psi}_2)$, where ${\cal S}_1\times {\cal S}_2$ are the s-sets of $(X_1 \times X_2)^T$ and $\hat{\Psi}_1 \otimes \hat{\Psi}_2: {\cal S}_1\times {\cal S}_2 \rightarrow {\cal H}_1 \otimes {\cal H}_2$ is defined by
\begin{equation}
\hat{\Psi}_1 \otimes \hat{\Psi}_2[(t, \Delta_1 \times \Delta_2)]:=\hat{\Psi}_1[(t, \Delta_1)] \otimes \hat{\Psi}_2[(t, \Delta_2)]
\end{equation}
on the s-sets of the type $(t, \Delta_1 \times \Delta_2)$, and then extended by $\sigma$-additivity to a generic s-set of ${\cal S}_1\times {\cal S}_2$.
%*****************************************************************
\section{Cournot's principle} \label{cournot}

Cournot's principle connects the mathematical formalism of probability theory to the empirical world. Let us consider an experiment represented by a suitable probability space. A possible formulation of the principle is the following:
\begin{trivlist}
\item[\hspace\labelsep{\bf Cournot's principle:}] a given event with probability very close to 1 will happen with empirical certainty in a single trial of the experiment.
\end{trivlist}
In this statement and in the following statements of the same kind, the word ``given'' means that the event is singled out in advance or at least independently with respect to the outcome of the experiment.

Cournot's principle has a very long history, which dates back to Bernoulli, even though not always assuming this name in the literature. In his {\it Ars Conjectandi} (1713) Bernoulli explains that we can treat very high probability as ``moral certainty'' \cite{shafer}. Cournot seems to be the first who explicitly recognized the role of this principle in connecting the mathematical formalism of probability with the empirical word \cite{cournot}. L{\'e}vy, a French mathematician, confirmed the thesis that Cournot's principle is the only way of connecting a probabilistic theory with the world outside mathematics \cite{shafer}. Borel defined it as ``the only law of chance'' \cite{borel}. Under the name of ``Principle B'', Cournot's principle was included by Kolmogorov in his formulation of probability theory \cite{kolmogorov}. In more recent years Cournot's principle, expressed in the language of typicality, has been used in various papers of Goldstein et al. in connection with Boltzmann's statistical mechanics \cite{goldstein} and Bohmian Mechanics \cite{durr1}. Rather surprisingly, in spite of this long history and for reasons that have never been clearly explained, Cournot's principle seems to have disappeared from most modern textbooks on probability. The history of the rise and fall of Cournot's principle can be found in various interesting papers of Shafer, for example in \cite{shafer}.

This section will discuss Cournot's principle and the reasons why it is necessary in order to derive empirical predictions from the mathematical formalism of probability theory. The reason for this preliminary section, as we will see, is that the new quantum principle proposed in this paper can be considered as the quantum version of Cournot's principle.

\vspace{3mm}
First of all, let us briefly review the way in which Cournot's principle, together with the product rule for independent trials, allows us to derive the well-known correlation between probability and relative frequency, that is the fact that in a long sequence of independent trials of an experiment the relative frequency of an event is approximately equal to its probability. Let $\cal E$ be a statistical experiment represented by a probability space ${\cal P}=(\Omega, {\cal F}, P)$. According to the product rule for independent trials, the experiment ${\cal E}_N$ whose trials are composed of $N$ independent trials of $\cal E$ is represented by the probability space ${\cal P}^N$. Given an event $A$ of $\Omega$, let us define on $\Omega^N$ the random variable
\begin{equation}
X_N[(\omega_1, \ldots, \omega_N)]:=\frac{\sum_{i=1}^N {\bf 1}_A(\omega_i)}{N},
\end{equation}
where ${\bf 1}_A$ is the characteristic function of the set $A$. The value of $X_N$ on a sequence $(\omega_1, \ldots, \omega_N)$ is the relative frequency of the event $A$ in the sequence. We can easily see that the expected value E$(X_N)=P(A)$ and the variance Var$(X_N)=P(A)P(A^c)/N$. From Chebyshev's inequality we obtain 
\begin{equation} \label{a}
P^N(|X_N - P(A)| \geq \epsilon) \leq \frac{P(A)P(A^c)}{\epsilon^2 N},
\end{equation}
where $\epsilon$ is a ``small'' positive number. Thus, for sufficiently large $N$, the event of $\Omega^N$ composed of those sequences for which the relative frequency of $A$ has a value close to $P(A)$ has a probability close to 1. Cournot's principle states that that event will happen.

From the inequality (\ref{a}) we obtain the limit
\begin{equation} \label{b}
\lim_{N \rightarrow \infty} P^N(|X_N - P(A)| \geq \epsilon) = 0,
\end{equation}
which is the weak law of large numbers. In many textbooks this law is considered as a proof of the correlation between probability and relative frequency, and no reference whatever is made to Cournot's principle. In other textbooks this correlation is derived from the strong law of large numbers. These conclusions are however not correct, because the limit (\ref{b}) alone does not predict how large $N$ must be in order to obtain a relative frequency of an event close to its probability; this prediction requires Chebyshev's inequality (\ref{a}) and Cournot's principle. For example, if $P(A)=.5$, from (\ref{a}) after $25000$ trials we obtain the relative frequency of $A$ as $.5 \pm .1$ with empirical certainty, i.e. with probability equal to $1- 10^{-3}$. Such a result is of course empirically confirmed.

For the same reason Cournot's principle cannot be replaced by a principle which directly correlates probability and frequency, such as the following: ``in a long sequence of trials, with empirical certainty the relative frequency of an event $A$ is approximately equal to $P(A)$''. Again this formulation does not provide any quantitative prediction about how long the sequence must be.

In the formulation of the principle the word ``certainty'' has been weakened by the attribute ``empirical''. This is necessary in order to prevent incorrect conclusions, such as: given any class $\{A_\sigma\}_{\sigma \in I}$ of events with probability close to 1, then the event $\cap_{\sigma \in I} A_\sigma$ will happen with certainty. This conclusion is obviously incorrect, because the event $\cap_{\sigma \in I} A_\sigma$ may be the empty set.

It is my belief that the fact that in many textbooks on probability theory Cournot's principle is not even mentioned derives from the fact that it appears to be obvious, almost a tautology. This is probably due to the fact that for all of us the term ``probability'' has an implicitly intuitive meaning from which it is difficult to distance ourselves. A useful strategy to avoid this psychological trap is to adopt a suggestion proposed by Wallace \cite{wallace} (although in a different context) and to use another name for probability, for example ``sorability''\footnote{Here a shorter name than ``sqwerdleflunkicity'', as proposed by Wallace, is used.}. Thus we can define a ``sorability'' space, with a ``sorability'' measure having the well-known mathematical properties, after which we can mathematically define conditional ``sorability'', and so on. Now it is easier to recognize that a ``sorability'' space is only a mathematical structure with no empirical meaning, and that an explicit connection with the empirical world is necessary. This connection is exactly that given by Cournot's principle.

\vspace{3mm}
Cournot's principle has two important features. The first of these is vagueness. It is in fact not possible to provide a precise value for the probability distinguishing the events that will happen with empirical certainty from the other events. It is not even possible to establish a precise quantitative correlation between empirical certainty and relative frequency, because we would obtain a circular definition. For example, in the formulation of Cournot's principle we cannot replace the sentence ``with empirical certainty'' with the sentence ``with empirical certainty, i.e. with on the average one exception every $[1 - P(A)]^{-1}$ trials'', because we would assume a correlation between probability and relative frequency in a postulate which is utilized to deduce just such a correlation.

The second important feature of Cournot's principle is that it makes no reference to the additive structure of the probability measure. This feature probably makes it easier for us to accept the quantum version of Cournot's principle, which will not be associated with an additive set function. The independence of Cournot's principle (expressed in terms of typicality) from the additive structure of the probability measure has also been pointed out by Goldstein \cite{goldstein}.

\vspace{3mm}
One last remark: There is a possible generalization of Cournot's principle, that could be named the {\it conditional Cournot principle}: let  $A$ and $B$ be two events such that 
\begin{equation} \label{g}
\frac{P(A \cap B)}{P(A)} \approx 1;
\end{equation}
then if $A$ happens, $B$ also happens with empirical certainty. Of course the (absolute) Cournot principle is a particular case of the conditional principle, but it is not clear whether the conditional principle can be derived from the absolute principle. However, given two generic events $A$ and $B$, the absolute principle is sufficient in order to deduce that, in a long sequence of trials, the frequency of $A \cap B$ relative to $A$ is approximately $P(A \cap B)/P(A)$.

\vspace{3mm}
Let us conclude this section by explicitly listing the three postulates connecting a probability space with the empirical world. The mathematical structure of a probability space is assumed to be already defined.
\begin{itemize}
\item[P1] {\bf Probability spaces and statistical experiments:} a probability space ${\cal P} = (\Omega, {\cal F}, P)$ is associated with every statistical experiment; the outcome of the experiment corresponds to an element of $\Omega$.
\item[P2] {\bf Product rule:} if ${\cal P}_1$ and ${\cal P}_2$  are the probability spaces associated with two independent experiments, then the probability space ${\cal P}_1 \times {\cal P}_2$ is associated with the compound experiment.
\item[P3] {\bf Cournot's principle: } a given event with probability very close to 1 will happen with empirical certainty in a single trial of the experiment.
\end{itemize}
In the next section three postulates very similar to these will be proposed in order to relate the mathematical formalism of quantum mechanics to the empirical world.
%*****************************************************************
\section{The quantum Cournot principle} \label{quantum}

The quantum version of Cournot's principle will be derived utilizing three subsequent statements which follow in a more or less inevitable way from a basic assumption.

Let us consider to the so called ``Einstein's Boxes'' thought experiment \cite{norsen}. At a time $t_I$ a particle is prepared in such a way that its wave function is subdivided into two separated impenetrable boxes. For example, we can put a particle in a box and then insert an impenetrable barrier in the box, dividing the box into two boxes and the wave function of the particle into two non-overlapping parts. The two boxes can also be transported to two very distant places. At a time $t_F$ the boxes are open in order to determine in which of these the particle is contained. Let us consider the following
 \begin{trivlist}
\item[\hspace\labelsep{\bf Basic assumption:}] the particle was in the box where it has been found at the time $t_F$ since the time $t_I$.
\end{trivlist}
According to the standard formulation of quantum mechanics this assumption is not justified and is in any case useless, since it is only a speculation about the value of an observable in the absence of a measurement. However, from the physical point of view the assumption is very reasonable, because the boxes are impenetrable and the particle cannot jump from one box to the other. Moreover, in Section \ref{retrodictions} we will see that an assumption of this type, even if relative to a macroscopic context, can in fact be considered as empirically verifiable. Let us therefore consider the basic assumption to be true and let us now study the possible consequences.

The first important consequence is that the wave function is no longer the most complete description of the state of a particle. In fact, according to the basic assumption, during the time interval $[t_I, t_F]$ the particle is inside one of the two boxes, but such information is not present in the wave function. The missing information relates to the position of the particle, and it is therefore natural to assume that a complete description of the particle includes its position at every time. This single assumption, without any further hypothesis for the trajectory, can be formalized and generalized as follows. Let $X$ be the configuration space of a quantum system, and let $T$ denote the time interval $[t_I, t_F]$. Then we can state that:
\begin{trivlist}

\item[\hspace\labelsep{\bf First statement :}] during the time interval $T$ the quantum system follows a trajectory of $X^T$.
\end{trivlist}
Note that neither continuity nor differentiability is required for the trajectories, because there is no reason to require these, at least in the non-relativistic domain. Thus, even if the particles have a definite position at every time, they do not have a definite momentum.

Let us now study another consequence of the basic assumption. It is reasonable to correlate the fact that the particle stays inside a box with the fact that the supports of the two parts of the wave function are disjoined. Let us therefore suppose that the wave function $\Psi(t)$ of a quantum system is the sum of two parts $\phi(t):=U(t)\phi_0$ and $\phi_\perp(t):=\Psi(t) - \phi(t)$ which are exactly non-overlapping at the times $t_1$ and $t_2$. If $\Delta_1$ and $\Delta_2$ are the supports of $\phi(t_1)$ and $\phi(t_2)$ respectively, they satisfy the condition
\begin{equation} \label{c}
U(t_2-t_1)E(\Delta_1)\Psi(t_1) = E(\Delta_2)\Psi(t_2).
\end{equation}
This reasoning can also be reversed; if $\Delta_1$ and $\Delta_2$ satisfy condition (\ref{c}), then $\phi(t):=U(t-t_1)E(\Delta_1)\Psi(t_1)=U(t-t_2)E(\Delta_2)\Psi(t_2)$ does not overlap $\phi_\perp(t)$ at the times $t_1$ and $t_2$. Using the set function $\hat{\Psi}$ defined by (\ref{yu}), condition (\ref{c}) becomes:
\begin{equation} \label{d}
||\hat{\Psi}(S_1) - \hat{\Psi}(S_2)||= 0,
\end{equation}
where $S_1$ and $S_2$ are the s-sets $(t_1, \Delta_1)$ and $(t_2, \Delta_2)$ respectively. We can therefore state that:
\begin{trivlist}
\item[\hspace\labelsep{\bf Second statement:}] let $S_1$ and $S_2$ be two given s-sets such that $||\hat{\Psi}(S_1) - \hat{\Psi}(S_2)||= 0$; then, if the trajectory of the system belongs to $S_1$ it also belongs to $S_2$ with certainty, and vice-versa.
\end{trivlist}
Recall that the word ``given'' means that the two s-sets are singled out in advance or at least independently from the actual trajectory of the system. It is also necessary to point out that the second statement does not exactly correspond to the basic assumption. In fact, the former does not require that the two parts of the wave function are non-overlapping during the time interval $[t_1, t_2]$, as the latter does. Thus it is more correct to say that the second statement is suggested by the basic assumption rather than derived from it.

The situation in which the wave function is split into parts with exactly disjoined supports does not reflect physical reality. More realistic is the situation in which the parts have almost exactly disjoined supports, and it is natural to generalize the second statement to cover this situation. In order to do this, the condition (\ref{d}) must be relaxed to the condition $||\hat{\Psi}(S_1) - \hat{\Psi}(S_2)|| \approx 0$. This requires the introduction of a normalization factor. Condition (\ref{d}) will therefore replaced by the following condition:
\begin{equation}
1 - \frac{||\hat{\Psi}(S_1) - \hat{\Psi}(S_2)||^2}{||\hat{\Psi}(S_1)||^2 + ||\hat{\Psi}(S_2)||^2} = \frac{2 Re \langle \hat{\Psi}(S_1)| \hat{\Psi}(S_2) \rangle }{||\hat{\Psi}(S_1)||^2 + ||\hat{\Psi}(S_2)||^2} \approx 1.
\end{equation}
The norms have been squared for reasons that will soon become clear. Note that 
$$
\frac{2 Re \langle \hat{\Psi}(S_1)| \hat{\Psi}(S_2) \rangle }{||\hat{\Psi}(S_1)||^2 + ||\hat{\Psi}(S_2)||^2} \leq 1 \; \; \hbox{and} \; \; \frac{2 Re \langle \hat{\Psi}(S_1)| \hat{\Psi}(S_2) \rangle }{||\hat{\Psi}(S_1)||^2 + ||\hat{\Psi}(S_2)||^2} = 1 \Leftrightarrow \hat{\Psi}(S_1) = \hat{\Psi}(S_2).
$$
Since the condition (\ref{d}) has been relaxed, the attribute ``with certainty'' must also be relaxed. We can therefore formulate the
\begin{trivlist}
\item[\hspace\labelsep{\bf Third statement (the quantum Cournot principle):}] let $S_1$ and $S_2$ be two given s-sets such that 
\begin{equation} \label{e}
\frac{2 Re \langle \hat{\Psi}(S_1)| \hat{\Psi}(S_2) \rangle }{||\hat{\Psi}(S_1)||^2 + ||\hat{\Psi}(S_2)||^2} \approx 1;
\end{equation}
then, if the trajectory of the system belongs to $S_1$ it also belongs to $S_2$ with {\it empirical} certainty, and vice-versa.
\end{trivlist}
There is a strong analogy between the quantum Cournot principle and the following probabilistic Cournot-like principle: let $A$ and $B$ be two given events of a probability space $(\Omega, {\cal F}, P)$ such that 
\begin{equation} \label{f}
\frac{2 P(A \cap B)}{P(A) + P(B)} \approx 1;
\end{equation}
then, if $A$ happens $B$ also happens with empirical certainty, and vice-versa.
Note for example that the expression (\ref{e}) reduces to 
\begin{equation}
\frac{ 2 P_t(\Delta_1 \cap \Delta_2)}{P_t(\Delta_1) + P_t(\Delta_2)} \approx 1
\end{equation}
when $S_1$ and $S_2$ are the equal-time s-sets $(t, \Delta_1)$ and $(t, \Delta_2)$, where $P_t$ denotes the probability measure $||E(\cdot)\Psi(t)||^2$ on $X$. The norms have been squared in (\ref{e}) in order to emphasize the analogy between (\ref{e}) and (\ref{f}). In the probabilistic case the principle can be derived from the conditional Cournot principle introduced in Section \ref{cournot}. This derives from the fact that 
\begin{equation}
\frac{2 P(A \cap B)}{P(A) + P(B)} \approx 1 \Leftrightarrow \frac{P(A \cap B)}{P(A)} \approx \frac{P(A \cap B)}{P(B)} \approx 1,
\end{equation}
as can easily be proven. On the contrary, a quantum condition of the type 
\begin{equation} \label{fac}
\frac{Re \langle \hat{\Psi}(S_1) | \hat{\Psi}(S_2) \rangle}{||\hat{\Psi}(S_2)||^2} \approx 1
\end{equation}
cannot be interpreted as the probabilistic condition (\ref{g}), and condition (\ref{e}) is irreducible. The reason will be explained in Section \ref{consistency}.

In my previous papers on this subject \cite{galvan1, galvan2, galvan3} the quantum Cournot principle was formulated with the name {\it quantum typicality rule}. In \cite{galvan3} the attempt was made to express the quantum typicality rule in terms of a set of probability measures. The aim was to remove the vagueness in the definition of the rule. Subsequently, I realized that, due to the presence of a vague principle in probability theory as well, namely Cournot's principle, defining the quantum typicality rule in terms of probability did not remove the vagueness. I also realized that in fact the quantum typicality rule itself could be considered as the quantum version of Cournot's principle. For these reasons a definition of the rule in terms of probability is renounced here, and another name is proposed for it.

\vspace{3mm}
The quantum Cournot principle admits an important particular case. By assuming that one of the two s-sets is equal to $X^T$, and considering the inequality
$$
||\hat{\Psi}(S)||^2 \leq \frac{2 ||\hat{\Psi}(S)||^2}{1 + ||\hat{\Psi}(S)||^2}=\frac{2 Re \langle \hat{\Psi}(X^T)|\hat{\Psi}(S) \rangle }{||\hat{\Psi}(X^T)||^2 + ||\hat{\Psi}(S)||^2},
$$
we obtain the following 
\begin{trivlist}
\item[\hspace\labelsep{\bf Particular case:}] if $S$ is a given s-set with $||\hat{\Psi}(S)||^2 \approx 1$, then the trajectory of the system will belong to $S$ with empirical certainty.
\end{trivlist}
Note that this particular case is very similar to the probabilistic Cournot principle. In Section \ref{statistical} we will see that, for this particular case, the quantum Cournot principle incorporates the Born rule.

\vspace{3mm}
Let us conclude this section by proposing a formulation of quantum mechanics which recalls as much as possible of the formulation of probability theory proposed in the previous section.

Recall that a canonical stochastic process is the triple $(X^T, \sigma({\cal S}), P)$, i.e. a probability space whose sample space is the space of all the possible trajectories over the configuration space $X$. A probability space whose sample space is a space of trajectories will be referred to as a {\it path space}, and it always corresponds to a stochastic process, as shown in \cite{galvan1}. In the same paper it has been also shown that path spaces can be used to represent the evolution of a wide range of different dynamical systems, either deterministic or stochastic. The way in which a path space acts as a dynamical law is simple: the dynamical system is assumed to follow a trajectory chosen at random from the associated path space. As in any probability space, the features of the trajectory are determined by the probability measure by means of Cournot's principle.

In section \ref{some} a quantum process was defined as the triple $(X^T, {\cal S}, \hat{\Psi})$. There is an obvious analogy between a quantum process and a canonical stochastic process $(X^T, \sigma({\cal S}), P)$, with the correspondences  $X^T \leftrightarrow X^T$, ${\cal S} \leftrightarrow \sigma({\cal S})$ and $\hat{\Psi} \leftrightarrow P$. Note for example that the set function $\hat{\Psi}$ satisfies two properties which resemble those of a probability measure: it is $\sigma$-additive and $||\hat{\Psi}(X^T)||^2=1$. There are however also differences between the two structures: $\cal S$ is not a $\sigma$-algebra and $\hat{\Psi}$ is vector valued.

By analogy with probability theory and with the interpretation of path spaces, we assume that a suitable quantum process is associated with every closed quantum system, possibly the universe. The quantum system is assumed to follow a trajectory of $X^T$, whose features are now determined by the set function $\hat{\Psi}$ by means of the quantum Cournot principle. A product rule analogous to the probabilistic product rule for independent experiments also exists in the quantum case: if ${\cal Q}_1$ and ${\cal Q}_2$ are the quantum processes associated with two closed quantum systems in the same time interval $T$, then the quantum process ${\cal Q}_1 \times {\cal Q}_2$ defined in section \ref{some} is the quantum process associated with the compound system.

Let us then summarize the new formulation of quantum mechanics in the following three postulates, which are analogous to the corresponding postulates for probability theory.
\begin{itemize}
\item[Q1] {\bf Quantum processes and quantum systems}: a quantum process ${\cal Q}= (X^T, {\cal S}, \hat{\Psi})$ is associated with every quantum system which is closed in the time interval $T$; the quantum system follows a trajectory of $X^T$.

\item[Q2] {\bf Product rule}: if ${\cal Q}_1$ and ${\cal Q}_2$ are the quantum processes associated with two closed quantum systems in the same time interval $T$, then the quantum process ${\cal Q}_1 \times {\cal Q}_2$ is associated with the compound system.

\item[Q3] {\bf Quantum Cournot principle}: let $S_1$ and $S_2$ be two given s-sets such that 
\begin{equation} \label{i}
\frac{2 Re \langle \hat{\Psi}(S_1)| \hat{\Psi}(S_2) \rangle }{||\hat{\Psi}(S_1)||^2 + ||\hat{\Psi}(S_2)||^2} \approx 1;
\end{equation}
then, if the trajectory of the system belongs to $S_1$ it also belongs to $S_2$ with empirical certainty, and vice-versa.
\end{itemize}
In the following, the above formulation of quantum mechanics will be referred to as ``the new formulation''. It is a trajectory based formulation, analogous to Bohmian mechanics or Nelson's stochastic mechanics \cite{nelson}. In this type of formulation the measurement process and the observers do not enter into the theory on a fundamental level. The way in which the standard quantum measurement theory with observables and self-adjoint operators can be derived in such formulations is discussed in Section \ref{statistical}. In Sections \ref{statistical} and \ref{quasi} we will see explicitly how well-known empirical facts, namely the results of the statistical experiments and the quasi-classical macroscopic evolution, can be explained in the new formulation.
%*******************************************
\section{On the consistency of the quantum Cournot principle} \label{consistency}
Before studying the predictive power of the quantum Cournot's principle it is useful to discuss its logical consistency. An example of a possible problem is the following: Suppose that there exist three s-sets $S$, $S_1$ and $S_2$, with $S_1 \cap S_2 = \emptyset$, such that $M_\Psi(S, S_1) \approx M_\Psi(S, S_2) \approx 1$, where the following notation has been introduced:
\begin{equation}
M_\Psi(S_1, S_2):=\frac{2 Re \langle \hat{\Psi}(S_1)| \hat{\Psi}(S_2) \rangle }{||\hat{\Psi}(S_1)||^2 + ||\hat{\Psi}(S_2)||^2}.
\end{equation}
It is obvious that in this case the quantum Cournot principle would not be consistent. The above situation cannot be realized \cite{galvan1}. In order to understand this in an intuitive manner, consider that the condition $M_\Psi(S_1, S_2) \approx 1$ corresponds to the condition $\hat \Psi(S_1) \approx \hat\Psi(S_2)$. Thus $\hat \Psi(S) \approx \hat\Psi(S_1)$ and $\hat \Psi(S) \approx \hat\Psi(S_2)$ implies $\hat \Psi(S_1) \approx \hat\Psi(S_2)$, from which $\langle \hat \Psi(S_1) | \hat \Psi(S_2) \rangle  \neq 0$.

This result however does not guarantee the consistency of the principle in all situations, and it would be desirable to find a general consistency criterion. A criterion is proposed here which is based on the notion of compatible stochastic process: a canonical stochastic process $(X^T, \sigma({\cal S}), P)$ is said to be {\it compatible} with a quantum process $(X^T, {\cal S}, \hat{\Psi})$ if 
\begin{equation} \label{h}
M_\Psi(S_1, S_2) \approx 1 \Rightarrow
M_P(S_1, S_2) \approx 1,
\end{equation}
where
\begin{equation}
M_P(S_1, S_2):=\frac{2 P(S_1 \cap S_2)}{P(S_1) + P(S_2)}.
\end{equation}
The consistency criterion we propose requires that the quantum process admits a compatible stochastic process. This criterion is vague, but of course also the quantum Cournot principle is vague. We are not able to provide a rigorous proof here of the existence in general of a compatible stochastic process.

\begin{figure}
\begin{center}
\includegraphics {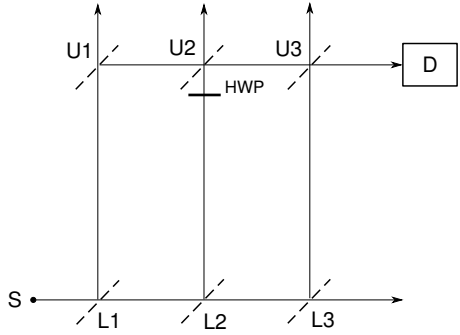}
\caption{a particle is emitted by the source S toward a series of beam splitters L1 ... U3. D is a detector. In the path L2-U2 there is an half-wave plate (HWP).} \label{fig1}
\end{center}
\end{figure}

\vspace{3mm}
Let us conclude this section by showing that the condition (\ref{i}) of the quantum Cournot principle is irreducible, as anticipated in section \ref{quantum}. The condition (\ref{i}) corresponds to the condition
\begin{equation}
E(\Delta_2)\Psi(t_2) \approx U(t_2 - t_1)E(\Delta_1)\Psi(t_1),
\end{equation}
from which we deduce that the value of $\Psi(t_2)$ in the region $\Delta_2$ depends almost totally on the value of $\Psi(t_1)$ in $\Delta_1$, and vice-versa. One could think that the first part of this property is also implied by the condition 
\begin{equation} \label{j}
E(\Delta_2)\Psi(t_2) \approx E(\Delta_2) U(t_2 - t_1)E(\Delta_1)\Psi(t_1),
\end{equation}
which however does not imply the reverse property. Thus we could think of associating this condition with the following property: if the trajectory of the system belongs to $S_2$, then it also belongs to $S_1$ with empirical certainty. This interpretation of (\ref{j}) is however inconsistent, because there are situations in which two disjoined sets $\Delta_1$ and $\Delta'_1$ exist such that 
\begin{equation}
E(\Delta_2)\Psi(t_2) \approx E(\Delta_2) U(t_2 - t_1)E(\Delta_1)\Psi(t_1) \approx E(\Delta_2) U(t_2 - t_1)E(\Delta'_1)\Psi(t_1).
\end{equation}
Consider for example the experiment shown in Fig. \ref{fig1}. Assume that the geometry of the apparatus is such that at a suitable time $t_M$ the wave function of the particle is split into three wave packets, whose supports are inside the three arms L1-U1, L2-U2 and L3-U3. Let $\Delta_1, \Delta_2$ and $\Delta_3$ be the supports of the three wave packets. Furthermore, let $t_F$ be the time at which the particle (possibly) reaches the detector D, and let $\Delta_F$ be the support of the corresponding wave packet. Due to the presence of the half-wave plate we have:
\begin{eqnarray}
E(\Delta_F)\Psi(t_F) & \approx & E(\Delta_F)U(t_F-t_M)E(\Delta_1)\Psi(t_M)  \nonumber \\ 
& \approx & - E(\Delta_F)U(t_F-t_M)E(\Delta_2)\Psi(t_M) \\
& \approx & E(\Delta_F)U(t_F-t_M)E(\Delta_3)\Psi(t_M). \nonumber
\end{eqnarray}
According to the tentative interpretation of (\ref{j}), we would then deduce that the particle travels along both the paths L1-U1 and L3-U3, which is of course impossible.

Note that the condition (\ref{j}) implies the condition (\ref{fac}). In fact the condition (\ref{fac}) is equivalent to the condition
$$
\frac{Re \langle \hat{\Psi}(S^c_1) | \hat{\Psi}(S_2) \rangle}{||\hat{\Psi}(S_2)||^2} \approx 0,
$$
and we have
\begin{eqnarray*}
& & \frac{Re \langle \hat{\Psi}(S^c_1) | \hat{\Psi}(S_2) \rangle}{||\hat{\Psi}(S_1)||^2} = \frac{Re \langle \Psi(t_2) | E(\Delta_2) U(t_2 - t_1)E(\Delta_1^c) | \Psi(t_1) \rangle}{||E(\Delta_2) \Psi(t_2)||^2} \leq \\
& & \frac{|| E(\Delta_2) U(t_2 - t_1)E(\Delta_1^c) \Psi(t_1)||}{||E(\Delta_2) \Psi(t_2)||} \approx 0.
\end{eqnarray*}
The second line of the above expression is the condition (\ref{j}) properly normalized. This proves the impossibility of interpreting (\ref{fac}) as the probabilistic condition (\ref{g}).
%*******************************************
\section{Statistical experiments and the Born rule} \label{statistical}

In the new formulation of quantum mechanics no mention has been made of the Born rule. In the context of a quantum process this rule could be formulated as follows:
\begin{trivlist}
\item[\hspace\labelsep{\bf The Born rule:}] the probability that the trajectory of the system belongs to a given s-set $S$ is $||\hat{\Psi}(S)||^2$.
\end{trivlist}
The Born rule has not been explicitly included in the new formulation because the quantum Cournot principle already incorporates its empirical predictions. Let us see how.

According to the Born rule, the experiment consisting of the measurement at the time $t$ of the position of a quantum system in the state $\Psi(t)$ is represented by the probability space $(X, {\cal B}, P_t)$, where $P_t:=||E(\cdot)\Psi(t)||^2$. As usual, in order to deduce that $P_t(\Delta)$ is the relative frequency with which $N$ systems in the same state $\Psi(t)$ are found in a region $\Delta$, it is necessary to consider the compound experiment, which is represented by the product space $(X^N, {\cal B}^N, P^N_t)$. Let $\Delta_{N, \epsilon}$ denote the event of $X^N$ composed by the sequences of outcomes $(x_1, \ldots, x_N)$ for which $\left |\sum_i  {\bf 1}_\Delta(x_i)/N -  P_t(\Delta) \right | \leq \epsilon$. With a reasoning analogous to that of section \ref{cournot} we deduce that for $N$ large enough the value of $P^N(\Delta_{N, \epsilon})$ is very close to 1. According to the probabilistic Cournot principle, the observed sequence will belong to $\Delta_{N, \epsilon}$.

Let us now apply the quantum Cournot principle. According to the new formulation, the quantum system is represented by the quantum process $(X^T, {\cal S}, \hat{\Psi})$, and the ensemble composed by $N$ copies of the system is represented by the quantum process $(X^{NT}, {\cal S}^N, \hat{\Psi}^N)$. It is easy to show that for any measurable set $\Sigma \subseteq X^N$ we have 
\begin{equation}
||\Psi^N[(t, \Sigma)]||^2 = P^N_t(\Sigma).
\end{equation}
Thus, for $N$ large enough the value of $||\Psi^N[(t, \Delta_{N, \epsilon})]||^2$ is very close to 1. By applying the (particular case of the) quantum Cournot principle we obtain the result that the trajectory of the ensemble will belong to $(t, \Delta_{N, \epsilon})$ with empirical certainty.

The conceptual structure of the two derivations can be represented in the following schematic ways. For the Born rule:
$$
\hbox{quantum formalism} \; \; \stackrel{BR}{\longrightarrow} \; \; \hbox{probability}\; \; \stackrel{CP}{\longrightarrow} \; \; \hbox{empirical predictions},
$$
where BR stands for the Born rule and CP for Cournot's principle. For the quantum Cournot principle:
$$
\hbox{quantum formalism} \; \; \stackrel{QCP}{\longrightarrow} \; \; \hbox{empirical predictions},
$$
where QCP stands for the quantum Cournot principle. These two schemes make evident the reason why it is unnecessary to explicitly include the Born rule in the new formulation of quantum mechanics. However, since we are used to thinking in terms of the Born rule and probability, in the rest of the paper we will assume the rule to be a part of the formulation, and use will be made of this in the discussion of some classical experiments.

\vspace{3mm}
In the above reasoning the equivalence between the Born rule and the quantum Cournot principle with regard to the prediction of the results of the statistical experiments has been proved only for position measurements. Moreover, in the new formulation the usual quantum measurement theory based on observables and operators is completely absent. Both these remarks are resolved by the fact that any measurement performed in a real laboratory ultimately comes down to a measurement of the position of a pointer, and the theory has a sufficient predictive power when it predicts the probability of the various positions of the pointer. The situation here is very similar to that in Bohmian Mechanics, and extensive theoretical work has been performed in order to derive an operator-based measurement theory from a position-based theory \cite{operators}. Here, we will briefly review just the way to associate a POVM (positive operator valued measure) with a measurement represented in a position-based approach.

Let our quantum system be a laboratory, that is a system composed of a microscopic system plus a measuring device. The measurement starts at the time $t_I$, at the end of the preparation phase, and ends at the time $t_F$, when the measuring device has recorded the outcome. At the time $t_I$ the state of the laboratory is $\phi \otimes \Phi$, where $\phi$ and $\Phi$ are the states of the microscopic system and of the measuring device, respectively. Let $\Omega$, equipped with some $\sigma$-algebra $\cal F$, be the set of the possible outcomes of the experiment. For example, in the measurement of the spin of a spin-$\frac{1}{2}$ particle along a suitable axis, $\Omega$ will be the set $\{+, - \}$. According to the position-based approach, a function $f:X \rightarrow \Omega$ exists that associates every configuration of the laboratory with the outcome corresponding to that configuration. Of course $f$ can be meaningfully defined only for a suitable subset of $X$, i.e. for those configurations corresponding to a measuring device which has recorded an outcome. We can however extend $f$ to the entire $X$ by adding a neutral element $\omega_0$ to $\Omega$ and associating this with the meaningless configurations. The function $f$ and the Born rule allow us to define the probability space $(\Omega, {\cal F}, P_\phi)$ representing the experiment, where
\begin{equation}
P_\phi(A):=||E[f^{-1}(A)] U(t_F-t_I) \phi \otimes \Phi ||^2.
\end{equation}
Note that the meaningless configurations will be outside the support of $U(t_F-t_I) \phi \otimes \Phi $, and therefore $P_\phi(\{\omega_0\}) \approx 0$. In order to associate a POVM with the measurement, for every event $A \in \cal F$ let us define the bilinear form
\begin{equation}
h_A(\phi, \varphi):= \langle \phi \otimes \Phi \mid U^\dagger (t_F-t_I)E[f^{-1}(A)] U(t_F-t_I) \mid \varphi \otimes \Phi \rangle,
\end{equation}
where $\phi$ and $\varphi$ are states of the Hilbert space of the microscopic system. We can easily see that $|h_A(\phi, \varphi)| \leq ||\phi|| ||\varphi||$, and therefore the form is bounded. Furthermore, it is self-adjoint and positive. From a theorem of functional analysis, it follows that a unique positive self-adjoint operator $O(A)$ exists such that $h_A(\phi, \varphi) = \langle \phi | O(A) | \varphi \rangle$. It is also easy to see that the mapping $O:A \mapsto O(A)$ is $\sigma$-additive. The mapping $O$ is therefore the POVM associated with the experiment, and we have $P_\phi(A)= \langle \phi \mid O(A) \mid \phi \rangle$.
%*******************************************
\section{Multiple time correlations} \label{multiple}

The fundamental feature distinguishing the quantum Cournot principle from the Born rule is that the former, but not the latter, establishes a correlation between the positions of a quantum system at different times. This is the key feature allowing the principle to define --under reasonable conditions for the wave function-- a dynamical structure for the trajectories and arguably explaining the observed quasi-classical evolution at the macroscopic level. In this section the nature of this correlation will be studied; the results will be used in the next section to prove that the trajectories have a quasi-classical structure.

Let us consider this first problem. Suppose we have a very large number $N$ of quantum systems in the same initial state. We recall that according to the new formulation, every system follows a well defined trajectory of $X^T$. Moreover, let us consider two s-sets $S_1$ and $S_2$, with $t_1 < t_2$, such that $M_\Psi(S_1, S_2)=1 - \epsilon$, with $\epsilon \ll 1$. Let $f_N(S_1), f_N(S_2), f_N(S_1 \cap S_2)$ denote the relative frequency of the events $S_1$, $S_2$ and $S_1 \cap S_2$ respectively. For instance, $f_N(S_1 \cap S_2)$ is $1/N$ times the number of systems whose trajectory belongs to $S_1 \cap S_2$. In the previous section we have seen that the quantum Cournot principle allows us to deduce the result
\begin{equation}
\lim_{N \rightarrow \infty} f_N(S_i)=||\hat{\Psi}(S_i)||^2 \; \; \hbox{for} \; \; i=1, 2.
\end{equation}
On the contrary, the quantum Cournot principle does not predict any value for 
\begin{equation} \label{wer}
\lim_{N \rightarrow \infty} f_N(S_1 \cap S_2).
\end{equation}
Note that the natural candidate expression for this value, namely $Re \langle \hat{\Psi}(S_1) | \hat{\Psi}(S_2) \rangle$, is not applicable, because it is not positive-definite. The incapacity of the quantum Cournot principle to predict the value of the limit (\ref{wer}) corresponds to the impossibility to experimentally determine {\it in principle} the value $f_N(S_1 \cap S_2)$. In fact the measurement of $\Delta_1$ at the time $t_1$ implies an interaction with the system at that time, thus violating the requirement that the system be closed during $T$. See the discussion in section \ref{retrodictions}, where it is shown that, even if the exact value of $f_N(S_1 \cap S_2)$ cannot be measured, the quantum Cournot principle can still be empirically verified by means of retrodictions, the (vague) way in which observers have access to the past configurations of the world.

In spite of the fact that no value for the limit (\ref{wer}) can be derived from the quantum Cournot principle, the following vague conclusion appears to be appropriate: if $M_\Psi(S_1, S_2) \approx 1$, the {\it overwhelming majority} of the trajectories belonging to $S_1$ also belong to $S_2$ with empirical certainty, and vice versa. Note that this statement could also be considered as an alternative formulation of the quantum Cournot principle. In the former formulation the vague notion of {\it overwhelming majority} replaces the vague notion of {\it empirical certainty} of the latter, but no precise correlation is established in the two formulations between the value of a quantum expression and the relative frequency of an event.

The above statement is just one example of a class of vague statements that we can reasonably expect to deduce from the quantum Cournot principle. In order to derive such statements in a standard way and for more complex situations, the following criterion is proposed: a suitable vague statement is valid for a quantum process according to the quantum Cournot principle if it is valid for all the stochastic processes compatible with the quantum process according to the probabilistic Cournot principle.

It is a straightforward matter to deduce the previous statement about the overwhelming majority of the trajectories by applying the proposed criterion. Let us than apply the criterion to verify the following statement: let $S_1, \ldots, S_N$ be $N$ given s-sets such that $||\hat{\Psi}(S_i)||^2 \approx 1$ for $i=1, \ldots, N$; then the trajectory of the system will belong with empirical certainty to the overwhelming majority of the $S_i$'s. Let us then attempt to derive such a result for a stochastic process $(X^T, \sigma({\cal S}), P)$ compatible with the quantum process. It is easy to see that the condition of compatibility (\ref{h}) implies that a positive number $\epsilon \ll 1$ exists such that  $P(S_i) \geq 1 - \epsilon $ for $i=1, \ldots, N$. Let us introduce the following random variable $Y:X^T \rightarrow [0,1]$:
\begin{equation}
Y(\lambda):=\frac{1}{N}\sum_{i=1}^N {\bf } {\bf 1}_{S_i}(\lambda).
\end{equation}
The value of $Y$ on the trajectory $\lambda$ is the fraction of $S_i$'s to which $\lambda$ belongs. It is easy to prove that E$(Y) \geq 1-\epsilon$ and
\begin{equation} \label{k}
P(Y \leq 1- \delta) \leq \frac{\epsilon}{\delta},
\end{equation}
where $\delta$ is a suitable ``small'' positive number. The inequality for the expectation value is straightforward. As to the inequality (\ref{k}), let $a$ be a given point of the interval $[0,1]$ and $0 \leq P_a \leq 1$ a given value (of probability). We then have 
\begin{equation}
\sup_{\{Y: P(Y \leq a )=P_a \} } \left \{\hbox{E}(Y) \right \} = a  P_a + (1 - P_a)=1 -P_a(1 - a).\end{equation}

In fact the supremum of the expectation value is obtained when the probability density $\rho(y)$ defined by $Y$ is shifted as far as possible to the right of the interval $[0,1]$, compatible with the constraint  $P(Y \leq a)=P_a$, that is when it is of the form $\rho(y)=\delta(y -a) P_a + \delta(y - 1) (1- P_a)$. The inequality (\ref{k}) is obtained by imposing the condition  $1 -P_a(1 - a) \geq 1 - \epsilon$ and by replacing $a$ with $1 - \delta$. If in (\ref{k}) we assume for example that $\epsilon = 10^{-9}$ and $\delta= 10^{-3}$ we obtain 
\begin{equation}
P(Y \leq 1 - 10^{-3}) \leq 10^{-6}.
\end{equation}
This inequality, together with the probabilistic Cournot principle, justifies the vague conclusion that the trajectory of the system will belong with empirical certainty to the overwhelming majority of the $S_i$'s. This conclusion is obviously valid for all the stochastic processes compatible with the quantum process and, according to the proposed criterion, can therefore be extended to the quantum process as well.

Equation (\ref{k}) establishes a quantitative correlation between the value of $\epsilon$ and the number of times the trajectory fails to belong to the $S_i$'s. As in the previous example, this quantitative correlation is lost when we consider the quantum process, for which only the vague conclusion that the trajectory belongs to the overwhelming majority of the $S_i$'s is valid. Here too this limitation is not a problem, because here too the number of s-sets to which the trajectory fails to belong cannot be empirically determined.

\section{The quasi-classical structure of the trajectories} \label{quasi}

Let us consider the general problem of proving that the trajectory of a quantum system has a quasi-classical structure at the macroscopic level. For this purpose, let us suppose that the wave function of the quantum system, possibly the universe, has the following structure: at the time $t_I$ it is a single wave packet; after a suitable time interval it splits into permanently non-overlapping wave packets; after another time interval some of the wave packets split again into permanently non-overlapping wave packets, and so on. For the time being let us assume this structure without further discussion; we will return to this assumption at end of the present section.

In order to mathematically represent such a structure let us define a {\it forward tree structure}, that is a set of $n$ mappings $\{\Delta_i:T \rightarrow {\cal B}\}_{i=1 \ldots n}$, satisfying suitable properties to be specified shortly. The principle is that the sets $\{\Delta_i(t)\}$ are the supports of the various wave packets at the time $t$. Any mapping $\Delta_i$ will be referred to as a {\it branch} of the tree structure.

First of all, the branches must satisfy the following conditions, required by the condition that they must form a forward tree-structure:
\begin{eqnarray}
& & \hbox{for} \; \;  i \neq j \; \; \hbox{we have either} \; \; \Delta_i(t) = \Delta_j(t) \; \; \hbox{or} \; \; \Delta_i(t) \cap \Delta_j(t) = \emptyset \; \; \hbox{for} \; \; t \in T; \\
& & \hbox{if} \; \; \Delta_i(t) = \Delta_j(t) \; \; \hbox{than} \; \; \Delta_i(s) = \Delta_j(s) \; \; \hbox{for} \; \; s \leq t;
\end{eqnarray}
The boundary conditions at $t_I$ and $t_F$ are:
\begin{eqnarray}
& & \Delta_i(t_I) = \Delta_j(t_I) \; \; \hbox{for any} \; \; i, j; \\
& & \Delta_i(t_F) \cap \Delta_j(t_F) = \emptyset \; \; \hbox{for} \; \; i \neq j.
\end{eqnarray}
For every $s \in T$, the index set $I=\{1, \ldots, n\}$ can be subdivided into disjoined subsets $\{I^s_1, \ldots I^s_{n_s}\}$ such that $\Delta_i(s)=\Delta_j(s)$ for $i, j$ belonging to the same index set $I^s_k$. For $s=t_I$ there is a single index set $I^{t_I}_1=I$. Let us associate the mappings $\Sigma^s_k: t \mapsto \Sigma^s_k(t):= \cup_{i \in I^s_k} \Delta_i(t)$ with any $I^s_k$. Expressed in words, $\Sigma_k^s$ is a branch that at the time $s$ ceases to split. Note that $\Sigma_1^{t_I}(t)$ is the support of $\Psi(t)$. Let $S^s_k(t)$ denote the s-set $(t, \Sigma^s_k(t))$.

The $\{\Delta_i\}$ are also required to satisfy the following conditions, corresponding to the requirement that the branches are the supports of permanently non-overlapping parts of the wave function:
\begin{eqnarray}
& & ||\Psi[S^{t_I}_1(t)]||^2 \approx 1 \; \; \hbox{for} \; \; t \in T; \\ \label{m} 
& & M_\Psi[S^t_k(s), S^t_k(t_F)] \approx 1 \; \; \hbox{for} \; \; t \leq s \in T \; \; \hbox{and} \; \; k=1, \ldots, n_t. \label{n}
\end{eqnarray}

Using a the criterion proposed in the previous section, let us prove that the trajectory of the system almost always remains inside a branch of the tree structure with empirical certainty. Let $(X^T, \sigma({\cal S}), P)$ be a stochastic process compatible with the quantum process representing our quantum system. It is easy to see that the compatibility condition (\ref{h}) implies that a positive number $\epsilon \ll1$ exists such that for any $t \in T$ we have
\begin{equation}
P[S^{t_I}(t)] \geq 1 - \epsilon / 2 \; \; \hbox{and} \; \; \frac{P[S^t_k(t) \cap S^t_k(t_F)]}{P[S^t_k(t)]} \geq 1 - \epsilon / 2.
\end{equation}
For $t \in T$ let us define the random variable $Y_t:X^T \rightarrow [0,1]$:
\begin{equation}
Y_t(\lambda)=\sum_{i=1}^n {\bf 1}_{\Delta_i(t_F)}[\lambda(t_F)] \cdot {\bf 1}_{\Delta_i(t)}[\lambda(t)].
\end{equation}
The value of $Y(\lambda)$ is 1 if $\lambda$ belongs to the same branch at the times $t$ and $t_F$, and 0 otherwise. We have E$(Y_t) \geq 1- \epsilon$. In fact
\begin{equation}
Y_t(\lambda)=\sum_{k=1}^{n_{t}} {\bf 1}_{S^{t}_k(t_F)}(\lambda) \cdot {\bf 1}_{S^{t}_k(t)}(\lambda).
\end{equation}
Thus,
\begin{equation}
\hbox{E}(Y_t)= \sum_{k=1}^{n_{t}} P[S^{t}_k(t_F) \cap S^{t}_k(t)] \geq 
(1- \epsilon / 2) \sum_{k=1}^{n_{t}} P[S^{t}_k(t)] = (1- \epsilon / 2) P[S^{t_I}_1(t)] \geq 1-\epsilon.
\end{equation}

Given any sequence of times $\{t_1, \ldots, t_N\}$ belonging to $T$, let us now define the random variable $Y:X^T \rightarrow [0,1]$:
\begin{equation} \label{o}
Y:=\frac{1}{N} \sum_{r=1}^N Y_{t_r}.
\end{equation}
The value of $Y$ on the trajectory $\lambda$ is the fraction of the times $\{t_1, \ldots, t_N\}$ for which the trajectory $\lambda$ belongs to the branch $\Delta_i$ such that $\lambda(t_F) \in \Delta_i(t_F)$. Since E$(Y_t) \geq 1- \epsilon$, then E$(Y) \geq 1 - \epsilon$ as well, and the inequality (\ref{k}) of the previous section holds true also for the random variable $Y$. According to our previous reasoning, the trajectory of the system will determine a value of $Y$ close to 1 with empirical certainty. Since the sequence $\{t_1, \ldots, t_N\}$ is arbitrary, we can therefore conclude that the trajectory of the system will belong to a branch of the tree structure with empirical certainty for the overwhelming majority of the times.

\vspace{3mm}
Various author assume that the wave function of a real macroscopic systems in fact admits a tree structure, for example \cite{bohm3, struyve, peruzzi, deotto}, even if they do not explicitly utilize the above mathematical definition of tree structure. The splitting of the branches corresponds to the measurement-like interactions. The fact that the branches are permanently non-overlapping depends on the irreversible interaction with the environment (decoherence). Recall that, due to the multidimensional structure of the wave function, two branches are non-overlapping if their supports differs just for the position of a single particle, for example a photon.

Note that the branches defined in this section are not ontological elements of the new formulation, but merely tools to approximately evaluate the trajectory of a quantum system. Thus, the fact that they are vaguely defined and that many different tree-structures can be defined for the same wave function is not a problem. The situation is different for the MWI, where the branches of the wave function, i.e. the worlds, appear to be the primitive ontology of that interpretation, and their vague definition is with certainty a problem.

Proving the existence of a tree-structure for the wave function of macroscopic systems solves just one half of the problem, because it is also necessary to show that the branches, or at least the overwhelming majority of them, are macroscopically localized and have a quasi-classical evolution. Arguably, this result can be derived from the Ehrenfest theorem and/or from a reasoning analogous to Mott's analysis of the bubble chamber \cite{mott}. The detailed study will not be developed here. 
%***********************************************************
\section{Analysis of some experiments} \label{analysis}
Let us examine some classical quantum experiments in the light of the new formulation. 

\begin{figure}
\begin{center}
\includegraphics {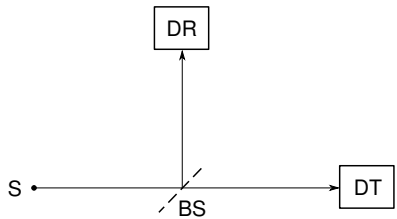}
\caption{A particle is emitted by the source S toward a beam splitter BS, and then detected by one of the two detectors DR or DT.} \label{fig2}
\end{center}
\end{figure}

\vspace{3mm}
{\it The beam splitter}. The first experiment is shown in Fig. \ref{fig2}. The quantum system here is the particle emitted by the source. The wave function admits an obvious tree structure with two branches: the branches coincide until the particle crosses the beam splitter. At this time the wave function splits into two non-overlapping wave packets which follow the two arms of the apparatus, and the two branches correspond to the supports of these two wave packets. According to the new interpretation the particle follows one of the two branches. This means that if the particle is detected, for example, by DR then it has travelled along the BM-DR arm. No physicist is likely to doubt this conclusion, even if it cannot be derived from standard quantum mechanics. Assumptions such as these are also made in the analysis of the so-called which-way experiments \cite{wheeler}. This experiment will be discussed again in Section \ref{retrodictions} in connection with the empirical verifiability of the quantum Cournot principle.

\begin{figure}
\begin{center}
\includegraphics {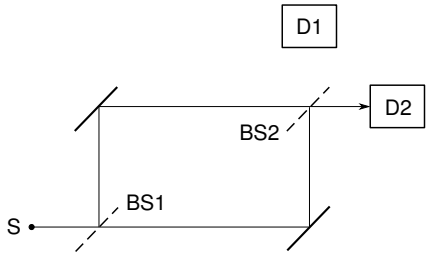}
\caption{A particle is emitted by the source S toward a Mach-Zender interferometer. Due to the interference the particle can reach only detector D2.} \label{fig3}
\end{center}
\end{figure}

\vspace{3mm}
{\it The double-slit experiment}. The experiment shown in Fig. \ref{fig3} corresponds to the double-slit experiment The wave function admits a tree structure composed by a single branch. However, every particle follows either the upper or the lower path between the beam splitters with probability $\frac{1}{2}$, and does not jump between the two paths. This can be deduced by applying the quantum Cournot principle to any one of the two parts into which the wave function is split between the two beam splitters. After BS2 the wave function is again a single wave packet, and the quantum Cournot principle does not allow us to retrodict the path followed by the particle. This means that the information of the path followed by the particle is lost for good.

Let us examine the reasons why this understanding of the experiment has not been accepted by the founding fathers of quantum mechanics and has led them to reject the possibility that particles follow definite trajectories \cite{heisenberg}. Suppose that the experimenter can freely close the lower path of the interferometer with a shutter. When the shutter is closed, the particles can also reach D1, which is forbidden when it is open. Thus, when a particle reaches BS2 through the upper arm, the particle and/or the beam splitter would have to ``know'' if the shutter is open or closed in order to know if the particle can be reflected towards D1 or not. But this is impossible, because the particle and the beam splitter are located well away from the shutter.

In my opinion, the above analysis is based on an implicit -- but possibly incorrect -- understanding that we have about random evolution, according to which it would have to be determined by some local stochastic interaction. However, this understanding is not even applicable to a classical stochastic process, i.e. to a system modelled as a set of paths endowed with a probability measure. In fact, the probability measure can favor some paths over others in a way which can be very different from the previous understanding, for example with global or teleological features. A quantum process has less structure than a stochastic process, but shares the same possibility to favor some sets of trajectories in a global way. Let us suppose that the shutter is controlled by a random mechanism, which makes its choice when a particle is emitted by the source, and let us include the shutter and the mechanism in the system represented by the quantum process. The wave function then admits a tree structure with four branches, corresponding to:  (1) shutter open, particle detected by D2; (2) shutter closed, particle detected by D1; (3) shutter closed, particle detected by D2;  (4) shutter closed, particle absorbed by the shutter. The quantum Cournot principle favors the trajectories belonging to these branches in a ``global'' way, and disfavors the other trajectories, for instance those corresponding to ``shutter open, particle detected by D1''.

\vspace{3mm} {\it Spin and the EPR paradox}. In a classical system represented by a stochastic process $(X^T, \sigma(\cal S), P)$ the trajectories of $X^T$ would have to be interpreted as the primitive ontology of the system, while the probability $P$ would have to be interpreted as the dynamical law governing the evolution of the system. By analogy, the elements $X^T$ and $\Psi$ of a quantum process would have to be interpreted as the corresponding elements of a stochastic process. For an N-particle system $X$ is equal to $ \RRR^{3N}$ and the spin variables are components of the wave function. As a consequence, the spin would have to be interpreted as a part of the dynamical law rather than a real physical property of the particles. The situation here is very similar to that in Bohmian mechanics \cite{operators}. 

The analysis of the Stern-Gerlach experiment is very simple: when the wave packet of a spin-$\frac{1}{2}$ particle crosses a Stern-Gerlach apparatus it splits (in general) into two non-overlapping wave packets. According to the new interpretation the trajectory of the particle will belong to the support of one of the wave packets, with a probability given by its squared norm.

Let us consider the EPR experiment: a spin-0 particle decays into two spin-$\frac{1}{2}$ particles, which travel toward two adjustable Stern-Gerlach apparatuses. Each apparatus can assume two different orientations. Let us assume that the two orientations are chosen by two independent random mechanisms, and let us include the apparatuses and the mechanisms in the system described by the quantum process. The wave function admits a tree structure with 16 branches, corresponding to four possible results for any one of the four possible orientations. The trajectory of the system will follow the support of one of these branches, with a probability given by the square norm of the branch.

What can we say about non-locality and the EPR-paradox? The fact is that here the spin of the particle is not considered a physical property of the particle, and is not therefore an element of reality. The reasoning leading to the EPR-paradox can thus not even begin.

\section{Retrodictions and the empirical verification of the quantum Cournot principle} \label{retrodictions}

It would appear that the quantum Cournot principle cannot be experimentally verified {\it in principle}. In fact, in order to test the principle we must measure the positions of a quantum system at two different times. If the evolution of the system is considered during the time interval $T=[t_I, t_F]$, we can choose $t_F$ as one of the two times, but the second time must be within the time interval $T$ (see note\footnote{The choice of $t_I$ as the first time does not allow a meaningful verification.}). This requires that an instrument interacts with the system at that time, thus violating the requirement that the system be closed during $T$.

Let us consider for example the experiment of the beam splitter described in Fig. \ref{fig2}, and let us suppose that we want to verify that a particle detected by DR has travelled along the BM-DR arm. In order to perform such a measurement, we can position a test particle at rest along the BS-DR arm and observe if it is scattered by the passage of the first particle. At least from the mathematical point of view, the mass of the test particle can be chosen as small as we want \footnote{The localization of the wave function of the test particle can be maintained by a suitable potential trap which does not influence the first particle.}, in such a way that the Hamiltonian of the first particle can be considered practically unaltered. Unfortunately, even if the test particle has a very small mass, the evolution of the first particle can be greatly influenced by the interaction with the test particle. In order to understand this, it is sufficient to apply such a measurement scheme to the interference experiment of Fig. \ref{fig3}: if we position a test particle at rest along one of the two paths between the beam splitters, the first particle can reach both detectors with the same probability, while in the absence of the test particle it can only reach D2.  We can therefore conclude that, in general, even in the presence of an apparently negligible interaction, the measured system can no longer be considered closed.

\vspace{3mm}
The above reasoning explain why we cannot experimentally determine the relative frequency $f_N(S_1 \cap S_2)$ of the intersection of two non-equal time s-sets (see section \ref{multiple}). In spite of this conclusion, the existence of retrodictions allows the quantum Cournot principle to be experimentally verified. The term {\it retrodiction} refers here to the possibility of an observer to obtain information about the configuration of a system at a time $t$ by performing a measurement at a time $t_F > t$ on the system itself or on another system which acts as a recording device, and therefore without disturbing the original system at $t$. Let us see in terms of an example how this can happen.

Let us consider again the experiment with the beam splitter of Fig. \ref{fig2}, but now let us also include the detectors in the system described by the quantum process. Let us single out the following times: at $t_I$ the particle has just been emitted by the source; at $t_1$ the wave function of the particle is inside the arms BS-DR and BS-DT; at $t_2$ the particle has just been detected by DR or DT; at $t_F$ the observer opens the door of the laboratory and determines which detector has been triggered. The entire laboratory (particle + detectors) can therefore be considered (at least ideally) as a closed system in the time interval $[t_I, t_F]$. The time $t_F$ can be much greater than the time $t_2$, and the detectors have of course been projected in such a way as to remain stable in the triggered state, i.e. to stably record the detection of the particle. The evolution of the wave function for the laboratory can then be represented as follows:
\begin{eqnarray}
& & \phi(t_I)  \otimes \Phi_R(t_I) \otimes \Phi_T(t_I) \rightarrow [\phi_R(t_1) + \phi_T(t_1)] \otimes \Phi_R(t_1) \otimes \Phi_T(t_1) \rightarrow  \\
& & \Phi_R^*(t_2) \otimes \Phi_T(t_2) + \Phi_R(t_2) \otimes \Phi_T^*(t_2) \rightarrow \Phi_R^*(t_F) \otimes \Phi_T(t_F) + \Phi_R(t_F) \otimes \Phi_T^*(t_F), \nonumber
\end{eqnarray}
where $\phi$ is the wave function of the particle and $\Phi_R$ and $\Phi_T$ are the wave functions of DR and DT, respectively. Moreover, $\phi_R$ and $\phi_T$ are the reflected and the transmitted wave packets of the particle, and $\Phi_R^*$ and $\Phi_T^*$ denote the triggered state of the detectors DR and DT, respectively. Note that 
\begin{eqnarray*} 
& & \Phi_R^*(t_F) \otimes \Phi_T(t_F) = U(t_F - t_2) \Phi_R^*(t_2) \otimes \Phi_T(t_2) = 
U(t_F - t_1) \phi_R(t_1) \otimes \Phi_R(t_1) \otimes \Phi_T(t_1); \\
& & \Phi_R(t_F) \otimes \Phi_T^*(t_F) = U(t_F - t_2) \Phi_R(t_2) \otimes \Phi_T^*(t_2) = 
U(t_F - t_1) \phi_T(t_1) \otimes \Phi_R(t_1) \otimes \Phi_T(t_1).
\end{eqnarray*}
Thus, if DR is triggered at $t_F$, i.e. if the configuration of the laboratory is inside the support of $\Phi_R^*(t_F) \otimes \Phi_T(t_F)$, the quantum Cournot principle states that it was inside the support of $\Phi_R^*(t_2) \otimes \Phi_T(t_2)$ at $t_2$ and inside the support of $\phi_R(t_1) \otimes \Phi_R(t_1) \otimes \Phi_T(t_1)$ at $t_1$.

The crucial point is then the following: When the observer opens the door and finds, for example, that DR is triggered he or she implicitly assumes that DR has detected the particle at the time $t_2$ and that it has remained in the triggered state until $t_F$, i.e. that the configuration of the laboratory has been inside the support of $U(t-t_2)\Phi_R^*(t_2) \otimes \Phi_T(t_2)$ for $t \in [t_2, t_F]$. Note that, from the formal point of view, this assumption is equivalent to the assumption that the particle detected by DR has travelled along the BM-DR arm, i.e. that at $t_1$ the configuration of the laboratory was inside the support of $\phi_R(t_1) \otimes \Phi_R(t_1) \otimes \Phi_T(t_1)$. Recall that, according to standard quantum mechanics, the latter assumption is not allowed and in any case useless, due to the lack of empirical evidence. In the case of the former assumption however, for the observer the intuitive belief that if DR is triggered at $t_F$ then the particle is detected at $t_2$ is so deeply rooted in his/her perception of reality that this can be considered empirical evidence. In other words, from the fact that the detector is triggered at $t_F$ the observer retrodicts the fact that it was triggered also at $t_2$, and for the observer these two facts represent the same degree of empirical evidence. This is the kind of empirical evidence which allows us to claim that, in this situation, the quantum Cournot principle is empirically verified.

\vspace{3mm}
It is easy to recognize that retrodictions are the basis of our empirical science and, more generally, of our perception of reality. Whenever we perform a measurement, for example the position of a planet, what we actually observe is the configuration of the instruments recording the position of the system at the time when such a position was measured. When we accept the fact that the system was actually there where the instruments say, we are making a retrodiction. A more basic example is when we observe an object: what we actually observe are the photons impinging on our retina; from this observation we retrodict where the object was a few $\mu$-seconds before.

There is a subtle point to be discussed relative to retrodictions. Observers retrodict the past configuration of a system $A$ on the basis of the present configuration of a system $B$, possibly the same system, which acts as a recording device. One could deduce that our perception of the past configuration of $A$ only depends on the present configuration of $B$, and that it is not necessary that $A$ was actually there where we retrodict it was. This is, for example, the position of Bell, who claims that ``we have no access to the past, but only to memories, and these memories are just part of the instantaneous configuration of the world''. This position leads to a very strange universe, in which there is no need of a dynamical law, and where the past does not correspond to our memories \cite{bell1, bell2, galvan1}. After proposing this kind of universe, Bell himself claims that it cannot be taken seriously. If the consequences of the reasoning of Bell cannot be taken seriously, then the reasoning must be incorrect. The point that Bell appears to miss is that the only information about the present configuration of $B$ is not sufficient to retrodict the past configuration of $A$, and that a dynamical law, i.e. a law correlating configurations at different times, is also necessary \cite{galvan1}. The fact that observers make retrodictions implies that a dynamical law does exist, and that in some way they have an intuitive knowledge of this law. 

Another remark. Usually we speak of knowledge of the present and memory of the past. Actually there is no true distinction between memory (retrodictions) and knowledge: memory can be considered as knowledge of the past, and the present is only the boundary of the past. Retrodictions (like knowledge) are imprecise and vague: Imprecise because observers cannot know the exact configuration of the universe, and vague because they cannot define a subset of the configuration space exactly describing their imprecise knowledge at a suitable time. Also the quantum Cournot principle defines the trajectories of a system in an imprecise and vague way. These common features makes it easier to accept an imprecise and vague law as a basic dynamical law, and even if the dynamical law were precise, it could not be tested empirically. This is the case for example with Bohmian mechanics, which defines exact trajectories for the particles, but these trajectories cannot be observed.

In Section \ref{quasi} we have seen that the trajectory of a system {\it almost} always remains inside a branch of the wave function. It is possible to observe when the trajectory goes outside the support of a branch? The answer is probably not, because retrodictions are based on the present configuration of our recording devices, and arguably they record only what happened with empirical certainty. In other words, sometimes the trajectory of the universe goes outside the support of a branch, but we cannot remember this.

%***********************************************************
\section{Summary} \label{summary}

Cournot's principle states that if a given event of a probabilistic experiment has probability close to 1 it will happen with empirical certainty in a single trial of the experiment. This vague principle is necessary to derive empirical predictions from the mathematical formalism of probability theory, namely to derive the well known correspondence between the probability and the relative frequency of an event.

A quantum version of the Cournot principle is proposed, based on the following assumption. Let us suppose that a quantum particle is prepared at a time $t_I$ in such a way that its wave function is split into two separate impenetrable boxes. At a time $t_F > t_I$ the boxes are open in order to check in what box the particle is contained. A natural assumption is that the particle has been in the box in which it has been found at $t_F$ since $t_I$. This assumption, which cannot be deduced from standard quantum mechanics, has two main consequences: (i) the wave function is not the most complete description of a particle, because during the time interval $[t_I, t_F]$ the particle is in one of the two boxes but such information is not present in the wave function; (ii) if the wave function of a particle is split into two non-overlapping parts, then  the trajectory of the particle stays inside the support of one of the two parts.

Point (ii) is the quantum Cournot principle. There is a strong analogy between the probabilistic and the quantum Cournot principles, and the two theories, probability theory and quantum mechanics, are formulated in such a way to emphasise a very similar conceptual structure.

According to the new formulation of quantum mechanics based on the quantum Cournot principle, the particles follows definite trajectories, as in Bohmian mechanics. The difference is that in Bohmian mechanics the trajectories are exactly defined by the guidance equation, while in the new formulation, under reasonable conditions for the wave function, they are approximately defined by the quantum Cournot principle.

Rather surprisingly, the principle also incorporates the empirical predictions of the Born rule, thus providing a unified explanation of the results of the statistical experiments and of the quasi-classical macroscopic evolution.

\end{document}